# Cherenkov like mechanism of surface wave excitation

V.S.Zuev, A.M.Leontovich, and V.V.Lidsky

The P.N.Lebedev Physical Institute of RAS
119991 Moscow, Russia
vizuev@sci.lebedev.ru
leon@sci.lebedev.ru
vlidsky@mail.ru

The theory is presented of surface wave excitation by a fast moving charged particle in a thin uniform metal film surrounded by a dielectic medium. It is shown that the Vavilov-Cherenkov like effect for the surface waves could arise for particles whose velocities are ten- and even hundred-times lower than the corresponding velocities in a uniform medium.

## Черенковский механизм возбуждения поверхностных волн

В.С.Зуев, А.М.Леонтович, В.В.Лидский

Физический ин-т им. П.Н.Лебедева РАН
119991 Москва
vizuev@sci.lebedev.ru
leon@sci.lebedev.ru
vlidsky@mail.ru

Предложена теория возбуждения поверхностных волн в тонкой однородной пленке металла, окруженной диэлектрической средой, при движении внутри пленки быстрой заряженной частицы. Показано, что эффект Вавилова-Черенкова для поверхностных волн возникает при скоростях частицы в десятки и сотни раз меньших, чем соответствующие скорости в однородной среде.

**Черенковский механизм возбуждения поверхностных волн**

В.С.Зуев, А.М.Леонтович, В.В.Лидский

Физический ин-т им. П.Н.Лебедева РАН
119991 Москва
vizuev@sci.lebedev.ru
leon@sci.lebedev.ru
vlidsky@mail.ru

1. Явление Вавилова-Черенкова – излучение света при движении быстрого электрона в среде наблюдают как в однородных средах /1,2/, так и в неоднородных средах, в таких, как фотонные кристаллы /3/. Условием возникновения этого излучения является наличие в пространстве (однородном или неоднородном) собственных электромагнитных волн с фазовой скоростью меньше, чем скорость пролетающего электрона.

Собственные волны с малой фазовой скоростью имеются в нанопленках и в нанонитях из серебра, золота, меди. Это так называемые поверхностные плазмон-поляритоны. Отличие фазовой скорости этих волн от скорости света в вакууме может составлять многие десятки и сотни раз. Это означает, что испускать излучение в виде плазмона может электрон, сравнительно медленный по сравнению с электроном, способным излучать в однородной среде /4/.

Неоднородное пространство с тонкой металлической пленкой имеет в качестве собственных волн симметричный и антисимметричный поверхностные плазмоны без продольной составляющей магнитного поля, так называемые $TM$ - плазмоны, плазмоны поперечно-магнитного типа с магнитным полем в плоскости пленки. Симметрию плазмонов мы определяем по виду магнитного поля. При заданном значении волнового числа частота симметричного плазмона выше частоты антисимметричного плазмона. Поле плазмонов локализовано на пленке, а на удалении от пленки экспоненциально мало. Таких волн в однородном пространстве нет.

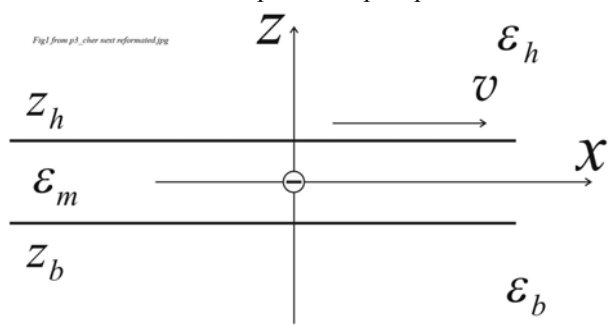

Рис.1.

2. Введем декартову систему координат, так чтобы частица двигалась вдоль оси $x$. Оси $x$ и $y$ лежат в плоскости пленки, ось $z$ направим перпендикулярно плоскости пленки, см. рис. 1. Ток, создаваемый точечной частицей, движущейся со скоростью $v$ вдоль оси $x$, и плотность заряда описываются следующими формулами:

$$j_x = ev\delta(x-vt)\delta(y)\delta(z), \; \rho = e\delta(x-vt)\delta(y)\delta(z). \qquad (2.1)$$

Решение уравнений Максвелла будем искать с помощью разложения полей на компоненты Фурье:

$$\vec{E} = \int_{-\infty}^{\infty} \vec{E}_\omega e^{-i\omega t} d\omega. \qquad (2.2)$$

Уравнения Максвелла для компоненты некоторой частоты $\omega$ принимают вид:

$$rot\vec{H}_\omega = -\varepsilon i\omega \vec{E}_\omega + \vec{v}\frac{2e}{v}\exp\left(i\frac{\omega x}{v}\right)\delta(y)\delta(z),$$

$$rot\vec{E}_\omega = i\omega\vec{H}_\omega, \qquad (2.3)$$

$$\varepsilon div\vec{E}_\omega = \frac{2e}{v}\exp\left(i\frac{\omega x}{v}\right)\delta(y)\delta(z).$$

Четвертое из уравнений Максвелла мы опустили, так как в данном случае оно автоматически вытекает из второго. Все компоненты поля зависят от координаты $x$ только благодаря фазовому множителю $\exp\left(i\frac{\omega x}{v}\right)$.

Проделаем теперь преобразование Фурье по волновым числам $k_y$:

$$\vec{E}_\omega = \exp(ik_x x)\int_{-\infty}^{\infty} \vec{E}_{\omega k_y}\exp(ik_y y)dk_y. \qquad (2.4)$$

В (2.4) мы ввели обозначение

$$k_x = \frac{\omega}{v} \qquad (2.5)$$

Для компонент полей $\vec{E}_{\omega k_y}$, $\vec{H}_{\omega k_y}$ из уравнений (2.3) получим следующую систему уравнений:

$$ik_y H_z - \partial_z H_y = -\varepsilon i\omega E_x + \frac{e}{\pi}\delta(z), \qquad (2.6)$$

$$\partial_z H_x - ik_x H_z = -\varepsilon i\omega E_y, \qquad (2.7)$$

$$ik_x H_y - ik_y H_x = -\varepsilon i\omega E_z, \qquad (2.8)$$

$$ik_y E_z - \partial_z E_y = i\omega H_x, \qquad (2.9)$$

$$\partial_z E_x - ik_x E_z = i\omega H_y, \qquad (2.10)$$

$$ik_x E_y - ik_y E_x = i\omega H_z. \qquad (2.11)$$

Для упрощения записи мы опустили у всех компонент индексы $\omega k_y$, указывающие по каким переменным сделано преобразование Фурье. Все величины в системе (2.6)-(2.11) зависят только от координаты $z$. Третье из уравнений (2.3) принимает вид:

$$\varepsilon(ik_x E_x + ik_y E_y + \partial_z E_z) = \frac{e}{\pi v}\delta(z). \qquad (2.12)$$

Несложно показать, что (2.12) является следствием уравнений (2.6)-(2.8).

3. Покажем, что система уравнений (2.6)-(2.11) распадается на две независимые системы уравнений. Введем в плоскости пленки новые координаты $\xi, \eta$ так, чтобы ось $\xi$ совпадала с направлением вектора $(k_x, k_y)$, а ось $\eta$ направим в плоскости пленки в перпендикулярном к оси $\xi$ направлении, при этом третья ось, ось $z$, (обозначение которой мы не меняем) сохраняет свое направление. Для компонент векторов напряженностей находим:

$$H_\xi = \frac{1}{k}(k_x H_x + k_y H_y) \qquad H_\eta = \frac{1}{k}(-k_y H_x + k_x H_y) \qquad (3.1)$$

$$E_\xi = \frac{1}{k}(k_x E_x + k_y E_y) \qquad E_\eta = \frac{1}{k}(-k_y E_x + k_x E_y) \qquad (3.2)$$

где через $k$ обозначено выражение

$$k = \sqrt{k_x^2 + k_y^2}. \qquad (3.3)$$

После такой замены неизвестных система (2.6)-(2.11) превращается в систему шести уравнений, которая распадается на две независимые тройки уравнений. Первая тройка уравнений:

$$\begin{cases} -\partial_z H_\eta + \varepsilon i\omega E_\xi = \frac{e}{\pi}\frac{k_x}{k}\delta(z) \\ k\cdot H_\eta + \varepsilon\omega E_z = 0 \\ \partial_z E_\xi - ik\cdot E_z - i\omega H_\eta = 0 \end{cases}. \qquad (3.4)$$

Вторая тройка:

$$\begin{cases} \partial_z H_\xi - ik\cdot H_z + \varepsilon i\omega E_\eta = -\frac{e}{\pi}\frac{k_y}{k}\delta(z) \\ \partial_z E_\eta + i\omega H_\xi = 0 \\ k\cdot E_\eta - \omega H_z = 0 \end{cases}. \qquad (3.5)$$

4. Рассмотрим последовательно системы уравнений (3.4), (3.5). Система уравнений (3.4) позволяет вычислить компоненты поля $H_\eta, E_\xi, E_z$. Три оставшиеся компоненты $H_\xi, H_z, E_\eta$ вычисляются совершенно независимо из системы уравнений (3.5). Для удобства нашего изложения мы сначала исследуем только решения системы (3.4) и определим компоненты $H_\eta, E_\xi, E_z$, а затем найдем оставшиеся компоненты полей из (3.5).

Исключив $E_z$ из второго и третьего уравнений (3.4), найдем:

$$H_\eta = \frac{\varepsilon i \omega}{k^2 - \varepsilon \omega^2} \partial_z E_\xi. \tag{4.1}$$

Подставив найденное $H_\eta$ в первое из уравнений (3.4), найдем:

$$\partial_z^2 E_\xi - (k^2 - \varepsilon \omega^2) E_\xi = -\frac{e}{\pi} \frac{k_x}{k} \frac{k^2 - \varepsilon \omega^2}{\varepsilon i \omega} \delta(z). \tag{4.2}$$

Это уравнение выполнено как внутри пленки, так и вне ее, причем для поля внутри под $\varepsilon$ следует понимать $\varepsilon_m$ — диэлектрическую проницаемость металла, а для поля выше и ниже пленки — соответственно $\varepsilon_h$ и $\varepsilon_b$.

Введем обозначения:

$$p = \sqrt{k^2 - \varepsilon_m \omega^2}, \quad q_h = \sqrt{k^2 - \varepsilon_h \omega^2}, \quad q_b = \sqrt{k^2 - \varepsilon_b \omega^2}. \tag{4.3}$$

Для полей в пленке и вне пленки получим уравнения:

$$\partial_z^2 E_\xi^{(m)} - p^2 E_\xi^{(m)} = -\frac{e}{\pi} \frac{k_x}{k} \frac{p^2}{\varepsilon_m i \omega} \delta(z), \tag{4.4}$$

$$\partial_z^2 E_\xi^{(h)} - q_h^2 E_\xi^{(h)} = 0, \tag{4.5}$$

$$\partial_z^2 E_\xi^{(b)} - q_b^2 E_\xi^{(b)} = 0. \tag{4.6}$$

5. Поскольку мы предполагаем $\varepsilon_m < 0$, ясно что значение $p$ (см. (4.3)) вещественно (мы временно не рассматриваем малую мнимую часть $p$, связанную с конечной проводимостью металла). Величины же $q_h, q_b$ могут быть как вещественны, так и мнимы. Сопоставляя (2.5), (3.3) и (4.3) мы видим, что рассматриваемое явление существенно зависит от соотношений между скоростью частицы и диэлектрической проницаемостью окружающей пленку среды.

Остановимся кратко на случае, когда скорость частицы превосходит величину фазовой скорости света в среде: $v > c/\sqrt{\varepsilon_h}$ (где $c$ — скорость света в вакууме). При малых значениях $k_y$ величина $q_h$ окажется чисто мнимой, следовательно решения уравнения (4.5) представляют собой осциллирующие функции $z$, что соответствует излучению электромагнитных волн в окружающее пленку пространство. Это классический случай черенковского излучения, теория которого предложена в /5/. В этом случае мнимую часть $q_h$ естественно обозначить $k_z$, так что $q_h = i k_z$. Излучение Вавилова-Черенкова сосредоточено на поверхности конуса, причем волновой вектор возникающих волн оказывается равным $\vec{k} = \{k_x, k_y, k_z\}$. Источником излучения является в нашем случае поле на поверхности пленки, индуцированное движущейся внутри металла частицей.

Картина качественно меняется в случае, когда скорость частицы меньше скорости света в среде: $v < c/\sqrt{\varepsilon_h}$. В этом случае классическое черенковское излучение отсутствует /5/, однако становится возможным излучение поверхностных волн. Поверхностная волна распространяется строго по поверхности пленки, в связи с чем ее волновой вектор принимает вид $\vec{k} = \{k_x, k_y, 0\}$.

6. Итак, в дальнейшем мы будем предполагать $v < c/\sqrt{\varepsilon_h}$. Общее решение уравнения (4.4) является суммой частного решения неоднородного уравнения и общего решения однородного. Частное решение находим методом вариации произвольной постоянной и добавляем общее решение, зависящее от двух постоянных:

$$E_\xi^{(m)} = \frac{e}{2\pi} \frac{k_x}{k} \frac{p}{\varepsilon_m i \omega} \left[ \theta(z) e^{-pz} + (1 - \theta(z)) e^{pz} \right] + C_1 e^{pz} + C_2 e^{-pz}. \tag{6.1}$$

Здесь $\theta(z)$ — функция Хевисайда. Решение уравнений (4.5) и (4.6) найдем из условия убывания поля на бесконечности:

$$E_\xi^{(h)} = C_3 \exp(-q_h z), \tag{6.2}$$

$$E_\xi^{(b)} = C_4 \exp(q_b z). \tag{6.3}$$

Здесь $C_1, C_2, C_3, C_4$ — константы, которые должны быть определены из граничных условий. Применив уравнение (4.1) ко всем трем средам найдем компоненты полей $H_\eta$:

$$H_\eta^{(m)} = \frac{e}{2\pi}\frac{k_x}{k}\left[-\theta(z)e^{-pz} + (1-\theta(z))e^{pz}\right] + \frac{\varepsilon_m i\omega}{p}\left[C_1 e^{pz} - C_2 e^{-pz}\right], \tag{6.4}$$

$$H_\eta^{(h)} = -\frac{\varepsilon_h i\omega}{q_h} C_3 \exp(-q_h z), \tag{6.5}$$

$$H_\eta^{(b)} = \frac{\varepsilon_b i\omega}{q_b} C_4 \exp(q_b z). \tag{6.6}$$

7. Граничные условия требуют непрерывности тангенциальных компонент $H_\eta$ и $E_\xi$ на поверхностях $z = z_h$ и $z = z_b$. Учитывая, что $z_h > 0$ и $z_b < 0$, находим:

$$\frac{e}{2\pi}\frac{k_x}{k}\frac{p}{\varepsilon_m i\omega}\exp(-pz_h) + C_1 \exp(pz_h) + C_2 \exp(-pz_h) = C_3 \exp(-q_h z_h) \tag{7.1}$$

$$\frac{e}{2\pi}\frac{k_x}{k}\exp(-pz_h) - \frac{\varepsilon_m i\omega}{p}\left[C_1 \exp(pz_h) - C_2 \exp(-pz_h)\right] = \frac{\varepsilon_h i\omega}{q_h} C_3 \exp(-q_h z_h) \tag{7.2}$$

$$\frac{e}{2\pi}\frac{k_x}{k}\frac{p}{\varepsilon_m i\omega}\exp(pz_b) + C_1 \exp(pz_b) + C_2 \exp(-pz_b) = C_4 \exp(q_b z_b) \tag{7.3}$$

$$-\frac{e}{2\pi}\frac{k_x}{k}\exp(pz_b) - \frac{\varepsilon_m i\omega}{p}\left[C_1 \exp(pz_b) - C_2 \exp(-pz_b)\right] = -\frac{\varepsilon_b i\omega}{q_b} C_4 \exp(q_b z_b) \tag{7.4}$$

Выражения (7.1-4) могут быть переписаны как система линейных уравнений с правой частью относительно четырех неизвестных констант $C_1, C_2, C_3, C_4$. Решение этой системы приводит к выражениям:

$$C_1 = \frac{e}{2\pi}\frac{k_x}{k}\frac{p}{i\omega\varepsilon_m}\frac{Q_b \exp(-2pz_b) + 1}{Q_b Q_h \exp(2pd) - 1}, \tag{7.5}$$

$$C_2 = \frac{e}{2\pi}\frac{k_x}{k}\frac{p}{i\omega\varepsilon_m}\frac{Q_h \exp(2pz_h) + 1}{Q_b Q_h \exp(2pd) - 1}, \tag{7.6}$$

где $d = z_h - z_b$ — толщина пленки, а через $Q_h, Q_b$ обозначены следующие выражения:

$$Q_h = \frac{\varepsilon_m q_h + \varepsilon_h p}{\varepsilon_m q_h - \varepsilon_h p} \qquad\qquad Q_b = \frac{\varepsilon_m q_b + \varepsilon_b p}{\varepsilon_m q_b - \varepsilon_b p} \tag{7.7}$$

Теперь мы можем вычислить компоненты Фурье составляющей напряженности $E_{z,\omega k_y}^{(h)}$ и $E_{z,\omega k_y}^{(b)}$ на поверхностях пленки. Используя (6.4) и второе из уравнений (3.4), после упрощений находим:

$$E_{z,\omega k_y}^{(h)}(z_h) = \frac{e}{2\pi}\frac{k_x}{\varepsilon_h \omega}\frac{Q_b \exp(-2pz_b) + 1}{Q_b Q_h \exp(2pd) - 1}(Q_h - 1)\exp(pz_h) \tag{7.8}$$

$$E_{z,\omega k_y}^{(b)}(z_b) = -\frac{e}{2\pi}\frac{k_x}{\varepsilon_b \omega}\frac{Q_h \exp(2pz_h) + 1}{Q_b Q_h \exp(2pd) - 1}(Q_b - 1)\exp(-pz_b) \tag{7.9}$$

В формулах (7.8) и (7.9) мы восстановили опущенные ранее индексы $\omega k_y$, указывающие по каким переменным сделано преобразование Фурье. Теперь мы можем определить зависимость от времени и пространственных координат составляющей электрического поля $E_z(x,y,z,t)$ на поверхности пленки, то есть при $z = z_h$. Для этого надо подставить разложение (2.4) в (2.2) и выполнить интегрирование:

$$E_z(x,y,z,t) = \int\limits_{-\infty}^{\infty} dk_y \int\limits_{-\infty}^{\infty} d\omega \cdot E_{z,\omega k_y} \cdot \exp\left(i\frac{\omega}{v}x - i\omega t\right) \cdot \exp(ik_y y). \tag{7.10}$$

8. Вполне аналогично тому, как были определены компоненты $E_{z,\omega k_y}$ на поверхностях пленки из системы уравнений (3.4), уравнения (3.5) позволяют определить компоненты $H_{z,\omega k_y}$. Мы не станем загромождать текст вычислениями и выпишем сразу окончательный ответ:

$$H_{z,\omega k_y}^{(h)}(z_h) = -i\frac{e}{\pi}\frac{k_y}{p - q_h} \cdot \frac{\exp(-2pz_b)\cdot\Omega_b + 1}{\exp(2pd)\cdot\Omega_h\cdot\Omega_b - 1} \cdot \exp(pz_h) \tag{8.1}$$

$$H_{z,\omega k_y}^{(b)}(z_b) = -i\frac{e}{\pi}\frac{k_y}{p - q_b} \cdot \frac{\exp(2pz_h)\cdot\Omega_h + 1}{\exp(2pd)\cdot\Omega_h\cdot\Omega_b - 1}\exp(-pz_b) \tag{8.2}$$

где через $\Omega_h, \Omega_b$ обозначены следующие выражения:

$$\Omega_h = \frac{q_h + p}{q_h - p} \qquad\qquad \Omega_b = \frac{q_b + p}{q_b - p} \tag{8.3}$$

9. Ниже приведены спектры, построенные по формуле (7.8). На рис.2a-b по горизонтальным осям отложены величины $k_x, k_y$ в обратных нанометрах. Частота волны связана с волновым числом соотношением $\omega = k_x \cdot v$. По вертикальной оси – компонента $E_z$ электрического поля на поверхности пленки $z = z_h$ вне пленки. Диэлектрическая проницаемость пленки металла вычисляется по формуле $\varepsilon_m(\omega) = 1 - \omega_{pl}^2/\omega^2 + 0.3i$, причем плазменная частота выбрана соответствующей длине волны в вакууме $\lambda = 150\,nm$. Диэлектрическая проницаемость пространства вне пленки $\varepsilon_b = \varepsilon_h = 1$. Скорость частицы выбрана $v = 0.22c$. Это на одну двадцатую превышает скорость резонансного плазмона на частоте, соответствующей длине волны в вакууме $\lambda = 800\,nm$. Толщина пленки металла составляет $d = 2\,нм$. Предполагалось, что частица движется вдоль поверхности $z = z_h$, то есть в (7.8) положено $z_h = 0, z_b = -d$.

На рис. 3a-b при тех же значениях параметров построен спектральные компоненты поля $H_z$. Мы видим, что амплитуды поля $H_z$ почти на три порядка меньше соответствующих амплитуд поля $E_z$. Физическая причина этого факта ясна. Как несложно показать, среди собственных волн тонкой пленки находятся только волны моды TM, в которых $H_z = 0$. Собственных волн с отличной от ноля $H_z$ в пленке нет. Мы видим, что поле $H_z$ имеет иную природу, чем поле $E_z$. Это не распространяющаяся по пленке волна, а поле индуцируемое током частицы. Таким образом, практически вся теряемая частицей энергия переходит в излучение волн TM моды.

Зависимость от пространственных координат $x, y$ составляющей напряженности электрического поля $E_z(x,y,z_h)$ на поверхности пленки была определена по формуле (7.10) численным интегрированием. Результаты численного расчета представлены на рис. 4a-b. Вычисления проводились при тех же значениях параметров среды и скорости частицы. При вычислении интегралов в (7.10) было выбрано значение $t = 0$. Частица находится в момент наблюдения в точке $x = 0, y = 0$. По осям отложены пространственные координаты в нанометрах. Мы видим, что излучение заполняет треугольник, являющийся двумерным аналогом трехмерного конуса из классического явления Вавилова-Черенкова.

10. Подведем итог.

Электрон, движущийся в тонкой пленке порождает поверхностные волны, причем механизм возбуждения поверхностных волн аналогичен механизму черенковского излучения. Возникает двумерный эффект Вавилова-Черенкова.

В отличие от электрона в однородной среде, рассматриваемый эффект могут порождать электроны, скорость которых в десятки и сотни раз меньше соответствующей скорости в однородной среде.

Возникающее излучение заполняет треугольник, являющийся двумерным аналогом черенковского конуса. Причем в отличие от классического эффекта Вавилова-Черенкова излучение не сосредоточено у поверхности конуса, а заполняет все пространство внутри треугольника. Естественно предположить, что эта особенность связана с сильной зависимостью от частоты диэлектрической проницаемости металла, в связи с чем возникает существенная зависимость от частоты фазовой скорости поверхностных волн.

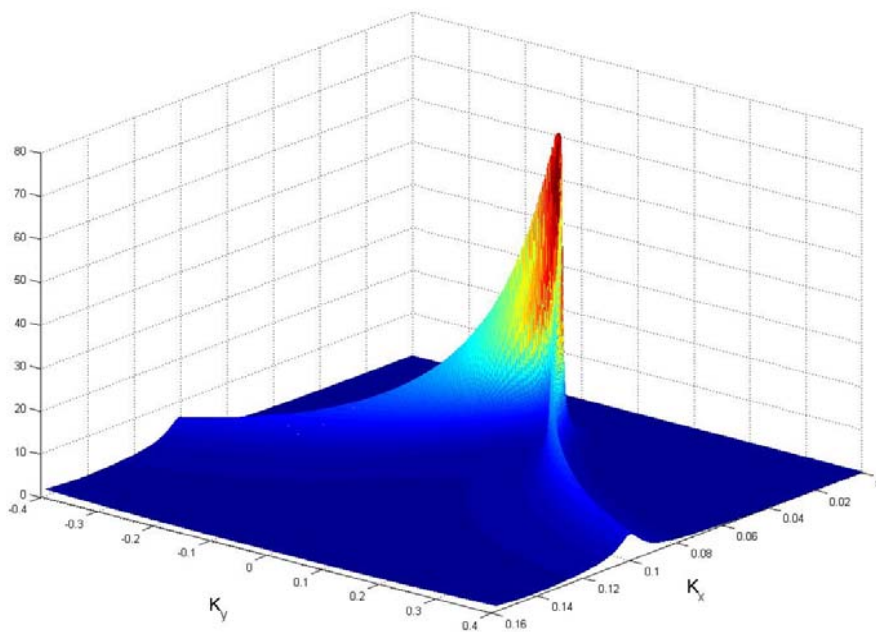

Рис. 2а. Спектр волн, возбуждаемых частицей, движущейся в пленке металла.

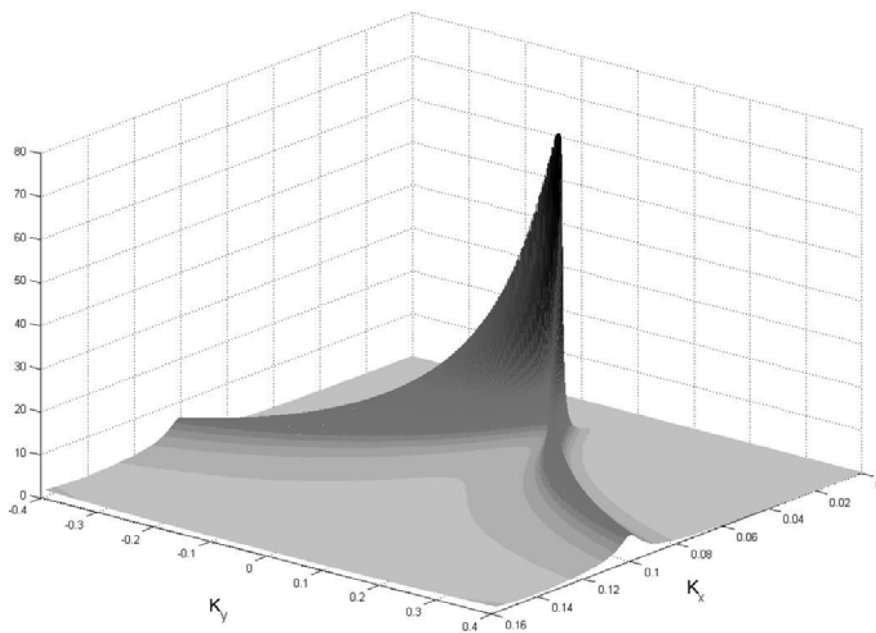

Рис. 2b. Спектр волн, возбуждаемых частицей, движущейся в пленке металла. Пояснения к графикам и параметры пленки помещены в тексте.

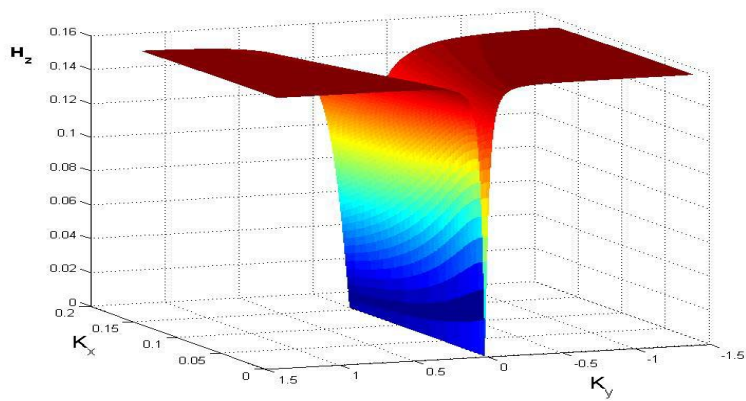

Рис. 3а.

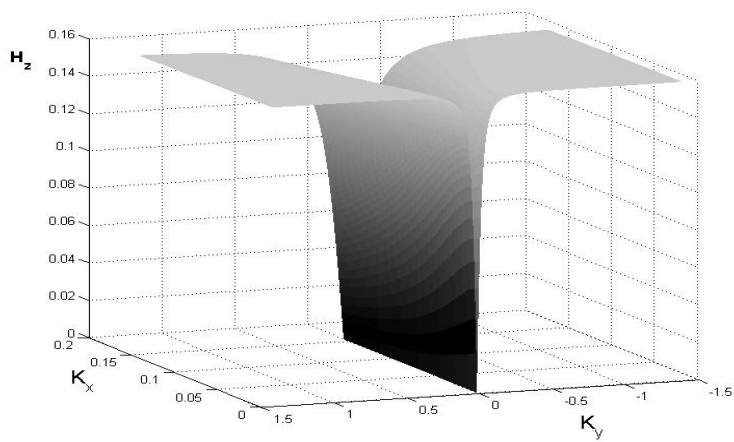

Рис. 3b. Зависимось от $k_x, k_y$ спектральной компоненты поля $H_z$. Параметры пленки, и единицы измерение напряженности такие же, как и при построении рис. 2.

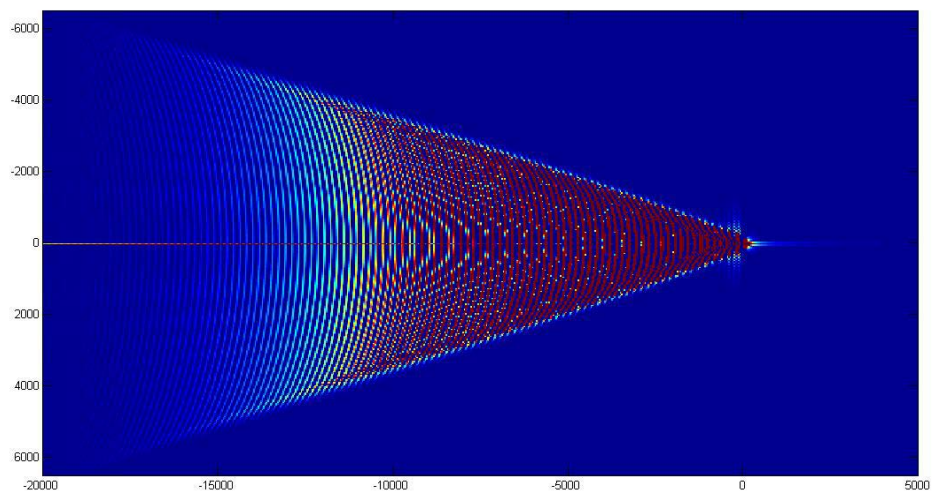

Рис. 4b

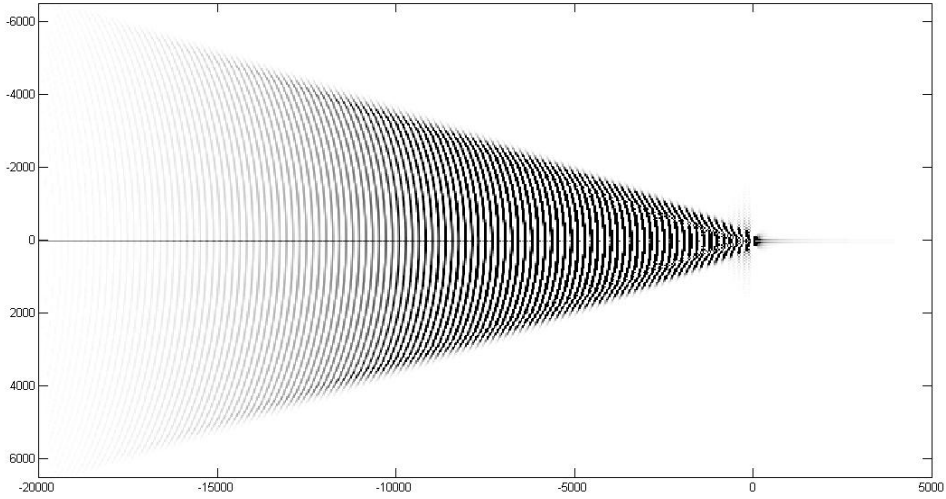

Рис. 4b. Зависимость от пространственных координат компоненты напряженности $E_z^{(h)}$ вблизи поверхности пленки. Частица движется вправо вдоль оси $x$, причем в момент наблюдения находится в точке $x = 0$. По осям отложены координаты в нанометрах. Параметры пленки приняты те же, что и при построении рис. 2.

1. П.А.Черенков. Труды ФИАН, т. 2, вып 4, стр 3-62 (1944)
2. Л.Д.Ландау, Е.М.Лифшиц. Электродинамика сплошных сред. Москва, Наука, 1982
3. C.Luo, M.Ibanescu, S.G.Johnson, J.D.Joannopoulos. Science, v.299, pp.368-371 (2003)
4. V.S.Zuev. Vavilov-Cerenkov phenomenon in metal nanofilms. arXiv:0907.1145 07 July 2009. В.С.Зуев. Оптика и спектроскопия, рег. №24709 от 22 июля 2009 г.
5. И.Е.Тамм, И.М.Франк, ДАН т.XIV(3),107-112(1937)

Приложения.

1. Если положить $\vec{H}_\omega, \vec{E}_\omega \sim \exp(i\omega x/v)$, то оказывается, что такое решение имеется у системы уравнений (2.3). Разложим поля $\vec{H}_\omega, \vec{E}_\omega$ по пространственным гармоникам вида $\vec{E}_\omega = \int\limits_{-\infty}^{\infty} \vec{E}_{\omega p} \exp(ipx)dp$ и подставим это выражение в уравнение. Тогда окажется, что для всех значений $p$, отличных от $\omega/v$ получим однородную систему уравнений относительно $\vec{E}_{\omega p}$. И лишь для $p = \omega/v$ получим уравнения с правой частью. Физически это означает, что решения с другими $p$ существовать могут, но они не имеют никакого отношения к рассматриваемой частице — их возбуждают какие-то другие источники.

2. О разложении δ-функции. Предположим $\delta(x - vt) = \int\limits_{-\infty}^{\infty} X_\omega e^{-i\omega t}d\omega$. Найдем $X_\omega$. Умножим обе части на $\exp(i\omega't)$ и проинтегрируем по $dt$ по отрезку $[-T, T]$.

$$\int\limits_{-T}^{T} \exp(i\omega't)\delta(x - vt) \cdot dt = \int\limits_{-\infty}^{\infty} X_\omega \int\limits_{-T}^{T} \exp(i(\omega'-\omega)t)dtd\omega.$$

Вычисляем интегралы по $dt$:

$$\frac{1}{v}\exp\left(i\frac{\omega'}{v}x\right) = \int\limits_{-\infty}^{\infty} X_\omega d\omega \frac{\exp(i(\omega'-\omega)T) - \exp(-i(\omega'-\omega)T)}{i(\omega'-\omega)}$$

или

$$\frac{1}{v}\exp\left(i\frac{\omega'}{v}x\right) = \int_{-\infty}^{\infty} X_\omega d\omega \frac{2\sin((\omega-\omega')T)}{(\omega-\omega')}$$

В интеграле в правой части делаем замену $\xi = (\omega - \omega')T$

$$\frac{1}{v}\exp\left(i\frac{\omega'}{v}x\right) = \int_{-\infty}^{\infty} X\left(\omega' + \frac{\xi}{T}\right) d\xi \frac{2\sin\xi}{\xi}$$

Это справедливо при любом $T$. Возьмем $T \to \infty$. Тогда ясно, что

$$\frac{1}{v}\exp\left(i\frac{\omega'}{v}x\right) = 2X(\omega')\int_{-\infty}^{\infty} d\xi \frac{\sin\xi}{\xi}$$

Оставшийся интеграл вычислим с помощью ТФКП. Подинтегральная функция регулярна в нуле, поэтому мы можем сместить контур несколько вниз и перейти от синуса к экпоненте:

$$\int_{-\infty}^{\infty} d\xi \frac{\sin\xi}{\xi} = \int_{-\infty-i\varepsilon}^{\infty-i\varepsilon} d\xi \frac{\sin\xi}{\xi} = \int_{-\infty-i\varepsilon}^{\infty-i\varepsilon} d\xi \frac{\exp(i\xi) - \exp(-i\xi)}{2i\xi} = \int_{-\infty-i\varepsilon}^{\infty-i\varepsilon} d\xi \frac{\exp(i\xi)}{2i\xi} - \int_{-\infty-i\varepsilon}^{\infty-i\varepsilon} d\xi \frac{\exp(-i\xi)}{2i\xi}$$

В первом из интегралов добавим путь по большим по модулю $\xi$ в верхней полуплоскости, замыкая путь от $+\infty$ к $-\infty$. А во втором интеграле замкнем контур в нижней полуплоскости. Первый интеграл содержит полюс $\xi = 0$ внутри контура интегрирования, а второй — не содержит. По теореме о вычетах

$$\int_{-\infty-i\varepsilon}^{\infty-i\varepsilon} d\xi \frac{\exp(i\xi)}{2i\xi} = \frac{2\pi i}{2i} = \pi; \quad \int_{-\infty-i\varepsilon}^{\infty-i\varepsilon} d\xi \frac{\exp(-i\xi)}{2i\xi} = 0$$

Таким образом

$$\int_{-\infty}^{\infty} d\xi \frac{\sin\xi}{\xi} = \pi, \ X(\omega') = \frac{1}{2\pi v}\exp\left(i\frac{\omega'}{v}x\right),$$

а значит

$$\delta(x - vt) = (1/2\pi v)\int_{-\infty}^{\infty} \exp(i\omega x/v)\exp(-i\omega t) d\omega.$$